\documentclass[11pt]{article}
\usepackage{hyperref}
\pdfoutput=1
\usepackage{natbib} % this is needed for JFM bibliography style
\usepackage[dvips]{graphicx}    % this is required to include eps files
\usepackage[dvips]{color}       % enables color
\usepackage{ulem} % required to xout text
\usepackage{amssymb,amsmath,amsfonts,amsbsy}
\usepackage{graphicx}
\usepackage{subfigure}
\usepackage{epsfig}
\usepackage{tabularx}
\usepackage[english]{babel}
\usepackage{wrapfig}
\usepackage{color}
\usepackage{supertabular}
\begin{document}
\title{LES of an inclined jet into a supersonic cross-flow}
\author{Antonino Ferrante$^1$, Carlos Pantano-Rubino$^2$, \\
Georgios Matheou$^1$, Paul E. Dimotakis$^1$,\\
Mike Stephens$^3$, Paul Adams$^3$, Richard Walters$^3$, Randall Hand$^3$
\\ \\
$^1$ Graduate Aeronautical Laboratories, \\ 
\vspace{6pt} California Institute of Technology, CA 91125, USA \\
$^2$ Mechanical Science and Engineering, \\
\vspace{6pt} University of Illinois at Urbana-Champaign, IL 61801, USA \\
$^3$ Data Analysis and Assessment Center, U.S. Army \\
Engineer Research and Development Center, MS 39180, USA}
\maketitle
%% The abstract (in this file, and that submitted as text to arXiv) should include the exact phrase
%% "fluid dynamics video" or "fluid dynamics videos"
\begin{abstract}
This short article describes flow parameters, numerical method, and animations of the fluid dynamics video \href{http://hdl.handle.net/1813/11480}{LES of an Inclined Jet into a Supersonic Cross-Flow}. Helium is injected through an inclined round jet into a supersonic air flow at Mach 3.6. The video shows 2D contours of Mach number and magnitude of density gradient, and 3D iso-surfaces of Helium mass-fraction and vortical structures. Large eddy simulation with the sub-grid scale (LES-SGS) stretched vortex model of turbulent and scalar transport captures the main flow features: bow shock, Mach disk, shear layers, counter-rotating vortices, and large-scale structures.
\end{abstract}
\subsection*{Flow description}
Helium is injected through an inclined round jet into a supersonic air flow (Fig.~1). In the present investigation, the jet axis forms a 30$^\circ$ angle with the streamwise direction of the air flow. The flow parameters of air and helium are reported in Table 1. The jet diameter, $d$, is 3.23$\times 10^{-3}$m, and the boundary layer thickness, $\delta$, of the air flow is 2$\times10^{-2}$m, as in the experimental study of \cite*{mad06}. The air free-stream Mach number is 3.6, the jet Mach number is 1.0, and the jet to free-stream momentum ratio, $q$, is 1.75. The Reynolds number of the air flow based on the momentum thickness is ${\sl Re}_\theta=U_{\rm e} \theta/\nu_{\rm w}=13 \times 10^3$ (${\sl Re}_\delta=U_{\rm e} \delta/\nu_{\rm w}=113\times10^3$), where $U_{\rm e}$ is the free-stream air velocity and $\nu_{\rm w}$ is the kinematic viscosity of air computed at the wall for adiabatic wall conditions.
\begin{figure*}[h]
\centering
\includegraphics[width=1\textwidth]{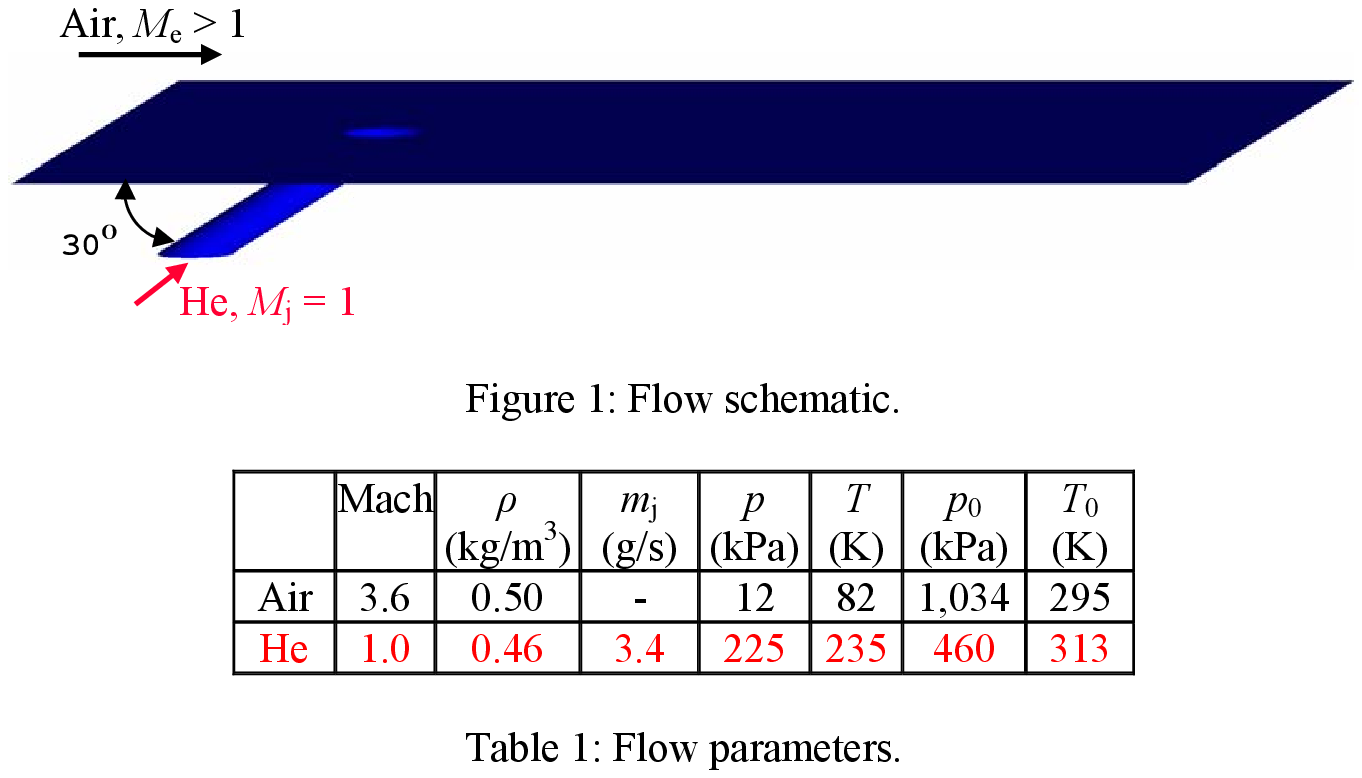} %%% NOTE: to make it work do not put extension .eps !!!
\vspace{-1cm}
\end{figure*}
\\ 
\subsection*{Numerical method}
We performed large-eddy simulation with the sub-grid scale (LES-SGS) stretched vortex model of turbulent and scalar transport developed by Pullin and co-workers \cite[]{pullin97,pullin00-1,pullin00,pullin02}. The governing equations are solved on a Cartesian mesh with adaptive mesh refinement (AMR) \cite[]{deiterd03}. 
The level-set approach with the ghost-fluid method \cite[]{fedkiw99} is used to treat the complex boundary (Fig.1) where no-slip and adiabatic boundary conditions are applied. 
The numerical method is a hybrid approach with low numerical dissipation \cite[]{hill04,pantano07} that uses tuned centered finite differences (TCD) \cite[]{hill04} in smooth flow regions, and weighted essentially non-oscillatory (WENO) \cite[]{osher94,shu96} scheme around discontinuities and ghost-fluid boundaries.
\newpage
\subsection*{Video}
The video \href{http://hdl.handle.net/1813/11480}{LES of an Inclined Jet into a Supersonic Cross-Flow} shows five animations:
\begin{enumerate}
\item Mach number contours in mid-span plane;
\item magnitude of density gradient contours in mid-span plane;
\item iso-surface of Helium mass fraction $Y_{He}=0.25$;
\item vortical structures\footnote{The vortical structures are educed using the $\lambda_2$-method \cite[]{hus95}, where $\lambda_2$ is defined as the second largest eigenvalue of the tensor $(S_{ik}S_{kj}+\Omega_{ik}\Omega_{kj})$, where $S_{ij}\equiv(\partial_jU_i+\partial_iU_j)/2$ is the strain rate tensor, and $\Omega_{ij}\equiv(\partial_jU_i-\partial_iU_j)/2$ is the rotation rate tensor.};
\item overlapped iso-surface of Helium mass fraction $Y_{He}=0.25$ and vortical structures.
\end{enumerate}
Each animation is played at a speed 10,000 times slower than in real life, and shows the flow evolution for about $1.8\times10^{-3}$~s. \\
The video shows that the main flow features are well captured: bow shock, Mach disk, shear layers, counter-rotating vortices, and large-scale structures \cite[]{af-etal-aps08}.
\subsection*{Acknowledgments}
This work was supported by AFOSR Grants FA9550-04-1-0020 and FA9550-04-1-0389, by the Caltech DoE Advanced Simulation and Computing (ASC) Alliance Center under subcontract No. B341492 of DOE contract W-7405-ENG-48, and NSF Grant EIA-0079871. 
The simulations were performed at the Center of Advanced Computing Research (CACR) at Caltech. 
The fluid dynamics video was produced at the Data Analysis and Assessment Center, U.S. Army Engineer Research and Development Center (ERDC).
%
%\bibliography{ref}

\begin{thebibliography}{0}
\expandafter\ifx\csname natexlab\endcsname\relax\def\natexlab#1{#1}\fi

\end{thebibliography}


\begin{thebibliography}{13}
\expandafter\ifx\csname natexlab\endcsname\relax\def\natexlab#1{#1}\fi

\bibitem[Deiterding(2003)]{deiterd03}
{\sc Deiterding, R.} 2003 {\em {Parallel Adaptive Simulation of
  Multi-dimensional Detonation Structures}\/}. Ph.D. Dissertation.

\bibitem[Fedkiw {\em et~al.\/}(1999)Fedkiw, Aslam, Merriman \& Osher]{fedkiw99}
{\sc Fedkiw, R.~P., Aslam, T., Merriman, B. \& Osher, S.} 1999 {A
  non-oscillatory {Eulerian} approach to interfaces in multimaterial flows (the
  {Ghost Fluid Method)}}. {\em J. Comput. Physics\/} {\bf 152}, 457--492.

\bibitem[Ferrante {\em et~al.\/}(2008)Ferrante, Pantano-Rubino, Matheou \&
  Dimotakis]{af-etal-aps08}
{\sc Ferrante, A., Pantano-Rubino, C., Matheou, G. \& Dimotakis, P.} 2008 {LES}
  of an inclined jet into a supersonic cross-flow at {Mach} 3.6. {\em Bull.
  {A}mer. {P}hys. {S}oc.\/} .

\bibitem[Hill \& Pullin(2004)]{hill04}
{\sc Hill, D.~J. \& Pullin, D.~I.} 2004 {Hybrid tuned center-difference-{WENO}
  method for large eddy simulations in the presence of strong shocks}. {\em J.
  Comput. Physics\/} {\bf 194}, 435--450.

\bibitem[Jeong \& Hussain(1995)]{hus95}
{\sc Jeong, J. \& Hussain, F.} 1995 On the identification of a vortex. {\em J.
  Fluid Mech.\/} {\bf 285}, 69--94.

\bibitem[Jiang \& Shu(1996)]{shu96}
{\sc Jiang, G.~S. \& Shu, C.~W.} 1996 {Efficient implementation of weighted ENO
  schemes}. {\em J. Comput. Physics\/} {\bf 126}, 202--228.

\bibitem[Kosovi{\'{c}} {\em et~al.\/}(2002)Kosovi{\'{c}}, Pullin \&
  Samtaney]{pullin02}
{\sc Kosovi{\'{c}}, B., Pullin, D.~I. \& Samtaney, R.} 2002 Subgrid-scale
  modeling for large-eddy simulations of compressible turbulence. {\em Phys.
  Fluids\/} {\bf 14}, 1511--1522.

\bibitem[Liu {\em et~al.\/}(1994)Liu, Osher \& Chan]{osher94}
{\sc Liu, X.~D., Osher, S. \& Chan, T.} 1994 Weighted essentially
  non-oscillatory schemes. {\em J. Comput. Physics\/} {\bf 115}, 200--212.

\bibitem[Maddalena {\em et~al.\/}(2006)Maddalena, Campioli \& Schetz]{mad06}
{\sc Maddalena, L., Campioli, T.~L. \& Schetz, J.~A.} 2006 Experimental and
  computational investigation of light-gas injectors in {Mach} 4.0 crossflow.
  {\em J. Propulsion and Power\/} {\bf 22}, 1027--1038.

\bibitem[Misra \& Pullin(1997)]{pullin97}
{\sc Misra, A. \& Pullin, D.~I.} 1997 A vortex-based subgrid stress model for
  large-eddy simulation. {\em Phys. Fluids\/} {\bf 9}, 2443--2454.

\bibitem[Pantano {\em et~al.\/}(2007)Pantano, Deiterding, Hill \&
  Pullin]{pantano07}
{\sc Pantano, C., Deiterding, R., Hill, D. \& Pullin, D.} 2007 {A low-numerical
  dissipation patchbased adaptive mesh refinement method}. {\em J. Comput.
  Physics\/} {\bf 221}, 63--87.

\bibitem[Pullin(2000)]{pullin00-1}
{\sc Pullin, D.~I.} 2000 A vortex-based model for the subgrid flux of a passive
  scalar. {\em Phys. Fluids\/} {\bf 12}, 2311--2319.

\bibitem[Voelkl {\em et~al.\/}(2000)Voelkl, Pullin \& Chan]{pullin00}
{\sc Voelkl, T., Pullin, D.~I. \& Chan, D.~C.} 2000 A physical-space version of
  the stretched-vortex subgrid-stress model for large-eddy simulation. {\em
  Phys. Fluids\/} {\bf 12}, 1810--1825.

\end{thebibliography}
\bibliographystyle{jfm}

\end{document}